\documentclass[%
 reprint,
 amsmath,amssymb,
 aps,
 pra,
 showkeys,
 floatfix,
]{revtex4-2}

\usepackage{multirow} 
\usepackage[utf8]{inputenc}
\usepackage[T1]{fontenc}
\usepackage{graphicx}
\usepackage{dcolumn}
\usepackage{bm}
\usepackage{booktabs}
\usepackage{array}
\usepackage[colorlinks=true, linkcolor=blue, citecolor=blue, urlcolor=blue]{hyperref}
\usepackage[capitalise]{cleveref}
\usepackage[all]{hypcap}
\usepackage{booktabs} 
\usepackage{amsmath}
\usepackage{amssymb}
\usepackage{mathtools}
\usepackage{mathrsfs}
\usepackage{amsthm}
\usepackage{amssymb}
\usepackage{url}
\usepackage{makecell}
\usepackage{xcolor,soul}
\sethlcolor{yellow}
\usepackage[ruled,vlined,linesnumbered]{algorithm2e}
\soulregister\cite7
\usepackage{float}

\begin{document}

\preprint{APS/123-QED}

\title{Quantum-inspired Ising machine using sparsified spin connectivity}
\author{Moe Shimada}
\author{Koki Awaya}
\author{Ryoya Yonemoto}
\author{Yu Zhao}
\author{Jun-ichi Shirakashi}
\email{shrakash@cc.tuat.ac.jp}
\affiliation{Department of Electrical Engineering and Computer Science, Tokyo University of Agriculture and Technology, Koganei, Tokyo, Japan.}

\date{\today}

\begin{abstract}
Combinatorial optimization problems become computationally intractable as these NP-hard problems scale. We previously proposed extraction-type majority voting logic (E-MVL), a quantum-inspired algorithm using digital logic circuits. E-MVL mimics the thermal spin dynamics of simulated annealing (SA) through controlled sparsification of spin interactions for efficient ground-state search. This study investigates the performance potential of E-MVL through systematic optimization and comprehensive benchmarking against SA. The target problem is the Sherrington–Kirkpatrick (SK) model with bimodal and Gaussian coupling distributions. Through equilibrium state analysis, we demonstrate that the sparsity control mechanism provides a consistent search of the solution space regardless of the problem's coupling distribution (bimodal, Gaussian) or size. E-MVL not only achieves the best performance among all tested algorithms—solving exact solutions up to 1600 spins where the best SA baseline is limited to 400 spins—but also provides insights that significantly improve SA's own temperature scheduling. These results establish E-MVL's dual contribution as both an efficient optimizer and a practical methodology for enhancing SA performance. Moreover, FPGA implementation achieved an approximately 6-fold faster solution speed than SA.
\end{abstract}


\maketitle
\raggedbottom

\section{Introduction}

Combinatorial optimization problems are NP-hard. Finding optimal solutions to them becomes computationally intractable as they scale, particularly in chemistry \cite{lin2025determination}, finance \cite{morapakula2025end}, and engineering \cite{sakai2019fabrication,kanezashi2025utility}. The computational complexity is exponential with increasing problem size \cite{siarry2016metaheuristics}, making large-scale optimization problems challenging. Various combinatorial optimization problems can be transformed into ground-state search problems of the Ising spin model \cite{barahona1982computational,lucas2014ising}. Therefore, developing efficient methods for solving the Ising spin model, particularly for spin‒glass problems, has become crucial for addressing these computational challenges. Several Ising spin computing methods have been developed, including quantum annealers \cite{morapakula2025end} and quantum-inspired Ising machines \cite{yamaoka201520k,takemoto20192,aramon2019physics,onizawa2025gpu,onizawa2024stochastic}. Compared with classical heuristic algorithms, quantum annealers with superconducting circuits offer faster ground-state searching \cite{denchev2016computational,albash2018demonstration}. However, the immature quantum technology and the high cost of systems operating at ultralow temperatures limit the broader adoption of quantum annealers. By contrast, quantum-inspired Ising machines with semiconductor circuits have been commercialized because of their stability, scalability, and low cost compared with quantum annealers.

Among the classical approaches to Ising computing, simulated annealing (SA) \cite{kirkpatrick1983optimization,isakov2015optimised} has been developed as a representative classical heuristic algorithm for solving optimization problems. However, recent studies have revealed that SA faces significant challenges when solving problems with Gaussian coupling distributions, particularly in spin-glass models, where SA algorithms demonstrate poor convergence properties compared with problems with bimodal couplings \cite{das2025classical,martinez2025problem}. The Sherrington–Kirkpatrick (SK) model \cite{sherrington1975solvable}, characterized by fully connected graphs with random couplings typically drawn from Gaussian or bimodal distributions, serves as a widely used benchmark for evaluating optimization algorithms \cite{aramon2019physics,das2025classical,martinez2025problem}. Although various research efforts have attempted to address these limitations through alternative temperature-based approaches \cite{das2025classical,martinez2025problem} and other advanced techniques, hardware-oriented approaches remain underexplored. In addition to SA, several advanced optimization algorithms have been applied to fully connected spin-glass problems. Population annealing combines sequential Monte Carlo resampling with SA and has been benchmarked on the SK model, although comparative studies have focused on relatively small system sizes ($N \leq 200$) \cite{martinez2025problem}. Momentum annealing incorporates momentum terms to accelerate escape from local minima and has been applied to fully connected problems with bimodal and Gaussian couplings equivalent to the SK model, with solution accuracy reported for systems up to approximately $N = 100$ \cite{okuyama2019binary}. Simulated quantum annealing introduces quantum tunneling effects through Suzuki–Trotter decomposition and represents an established quantum-inspired approach \cite{okuyama2017ising}. However, comprehensive benchmarking of these methods at larger scales remains limited.

Spin decision logic (SDL), which was initially proposed and implemented in HITACHI’s CMOS annealing \cite{yamaoka201520k,takemoto20192}, uses digital logic circuits for ground-state searches \cite{ito2017prompt,shimada2019calculation,miki2021hybridization,yoshida2022efficient,yoshida2022mimicking} In the SDL implementation, spin states $\{-1, +1\}$ can be efficiently mapped to digital bits $\{0, 1\}$, and exchange interactions can be represented using exclusive-NOR (XNOR) gates \cite{yamaoka201520k,takemoto20192} Compared with SA, this approach can be implemented in hardware with reduced resources and power consumption. We previously proposed extraction-type majority voting logic (E-MVL) \cite{yoshida2022efficient,yoshida2022mimicking}, which is an SDL; a sparse operation replaces the thermal fluctuations in SA. In our previous work, we demonstrated that E-MVL can effectively sample the equilibrium states of the Ising spin model close to the Markov chain Monte Carlo (MCMC) simulation results. Moreover, E-MVL may have better properties than SA in reaching the ground state, but systematic performance evaluations of established SA methods are lacking \cite{yoshida2022mimicking}.
This study addresses the comprehensive optimization of E-MVL parameters and provides a systematic performance evaluation against SA algorithms across different problem complexities and coupling distributions. A key methodological contribution is our equilibrium state analysis, which validates the theoretical foundation of E-MVL in statistical mechanics and reveals important insights into the relationship between the sparsity parameter of E-MVL and the temperature parameter of SA across different coupling distributions and sizes. We optimize the sparsity scheduling and iteration parameters of E-MVL to maximize both the solution accuracy and computational efficiency. We also demonstrate the performance of E-MVL via step-to-target (STT) \cite{kanao2022simulated,kuroki2024classical} and step-to-solution (STS) \cite{kanao2022simulated,kuroki2024classical} metrics, which evaluate intrinsic computational requirements independent of implementation factors. In this work, we demonstrate that the sparsity-based approach of E-MVL enables consistent energy landscape exploration across different coupling distributions (SK-bimodal and SK-Gaussian), establishing the statistical mechanical foundation of the algorithm. Through experimental validation demonstrating E-MVL’s consistent performance across different coupling distributions in SK models for systems up to 1600 spins, particular advantages are observed for challenging problem instances where traditional SA methods face computational limitations. The validation of the hardware implementation on FPGA demonstrates the computational efficiency advantages of E-MVL in dedicated hardware platforms.

\section{Methods}

\subsection{Convergence behavior in Ising model and simulated annealing}

\begin{figure*}[t]
\centering
\includegraphics[width=\textwidth]{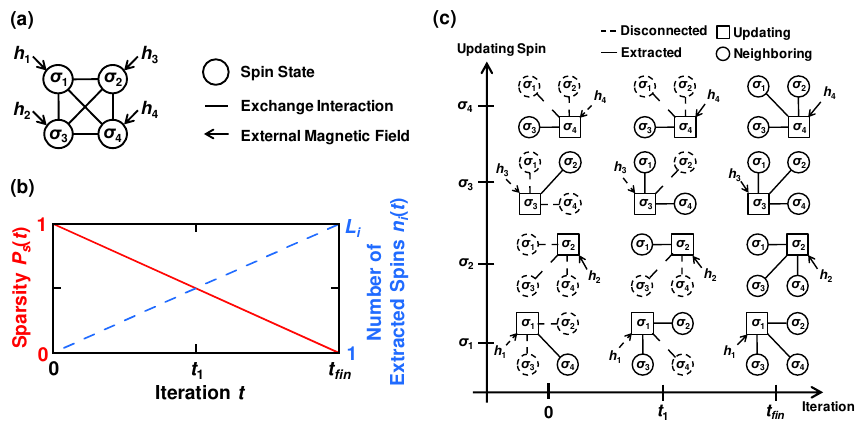}
\caption{Ising model and sparsification mechanism of E-MVL. (a) 4-spin fully connected Ising model. (b) Evolution of interaction sparsification corresponding to sparsity $P_{s}(t)$ reduction during ground-state search. As the number of iterations increases, the sparsity parameter $P_{s}(t)$ (solid line) decreases, whereas the number of extracted spins $n_{i}(t)$ (dashed line) increases. (c) Temporal evolution of spin-state updates with controlled sparsification of interactions following the schedule shown in (b). Three distinct phases of the spin update process show how the number of extracted spins increases as the sparsity decreases.}
\label{fig:Figure1}
\end{figure*}

The Ising spin model is a theoretical model that describes the behavior of magnetic spins \cite{brush1967history}. Fig.~\ref{fig:Figure1}(c) shows a 4-spin fully connected Ising model as an illustrative example. The Hamiltonian of the model is expressed as

\vspace*{-12pt}
\begin{equation}
H = - \sum_{i, j} J_{ij} \sigma_i \sigma_j - \sum_i h_i \sigma_i,
\label{eq:hamiltonian}
\end{equation}

where $\sigma_i \in \{+1, -1\}$ represents the spin state, $J_{\mathit{ij}}$ is the exchange interaction coefficient between $\sigma_{\mathit{i}}$ and $\sigma_{\mathit{j}}$, and $h_{\mathit{i}}$is the external magnetic field coefficient. Various combinatorial optimization problems can be transformed into ground-state search problems of the Ising model and solved via natural convergence properties. However, local minima in the energy profiles degrade the solution accuracy. Thermal fluctuations in SA can prevent the spin configuration from reaching local minima. SA’s convergence theorem guarantees solution quality because the spin states are iteratively updated to reach the Boltzmann distribution \cite{kuroki2024classical}.

\subsection{Previous approaches to spin decision logic}
SDL can perform energy convergence calculations via logic circuits by representing the spin states $\{-1, +1\}$ as digital bits $\{0, 1\}$. By using spin state $\sigma_{\mathit{j}}$ and interaction coefficient $J_{\mathit{ij}}$ as inputs to XNOR gates, the output of the XNOR gates becomes the next spin state of $\sigma_{\mathit{i}}$ that reduces energy. Because this eliminates the need for numerical calculations, it enables hardware implementation with reduced resources and power consumption compared with SA. Using only XNOR gates results in convergence to the nearest local minimum from the initial spin state. Therefore, each SDL implements a different method to escape local solutions. In majority voting logic (MVL) \cite{yamaoka201520k,takemoto20192}, which was first proposed as an SDL and utilized in HITACHI’s CMOS annealing implementation, the spin state is updated to a lower-energy state by majority voting circuits. Inversion circuits are used to flip the output of the majority voting circuits to avoid local minima. Similar to SA, which can search for the ground state on the basis of thermal fluctuations and temperature scheduling, random fluctuations and spin-flip probability control lead to energy convergence. However, random fluctuations do not guarantee that sufficiently slow energy convergence will ensure a ground state, contrary to convergence \cite{granville1994simulated} or quantum adiabatic theorems \cite{albash2018adiabatic}. Several modified MVLs have been proposed to achieve better performance. Prompt decision logic (PDL) \cite{ito2017prompt,shimada2019calculation} randomly extracts 1 connected spin and cuts the interactions between $\sigma_{\mathit{i}}$ and other connected spins when updating the target spin $\sigma_{\mathit{i}}$. This process stochastically transitions the spin states to higher energies and avoids local minima. Moreover, PDL reaches the ground state by repeating these processes for an extensive period, resulting in higher solution accuracy than MVL in the maximum-cut problem \cite{ito2017prompt,shimada2019calculation}. In addition, PDL is unnecessary for setting inversion circuits and scheduling fluctuations. However, PDL cannot be applied to Ising problems with all-to-all connectivity because it consistently references only a single spin from all connected spins during spin updates, making it impossible to control the fluctuations necessary for ground-state convergence.

\subsection{Extraction-type majority voting logic}

E-MVL is a type of SDL that introduces a novel approach to ground-state search through controlled sparsification of spin interactions. Although the basic E-MVL algorithm was previously introduced \cite{yoshida2022mimicking}, this study presents significant improvements in parameter optimization and performance evaluation. The core principle of E-MVL lies in the efficient search for the ground state of the Ising model through sparsification of the spin interactions. Here, sparsification refers to temporarily disconnecting some of the interactions among connected spins and selectively considering only a subset of the remaining connected spins during the spin-state updates. The parameter that controls the disconnection rate of the connected spins is called sparsity $P_{s}(t)$ \cite{yoshida2022efficient, yoshida2022mimicking}. $P_{s}(t) =$ 0 means that all interactions are connected (no sparsification), and $P_{s}(t) =$ 1 means that all interactions are disconnected. The sparsity $P_{s}(t)$ takes values ranging from 0 to 1, indicating the proportion of disconnected spin interactions at iteration t. Typically, $P_{s}(t)$ starts with a value close to 1 (disconnecting many spin interactions) and gradually decreases toward 0 (considering more connected spins) as the iterations progress. This transition facilitates a gradual shift from initial global exploration to final energy convergence.

By introducing the sparsity parameter $P_{s}(t)$, E-MVL can control the fluctuations in the PDL. Whereas PDL consistently extracts only 1 spin ($n_{i}(t) =$ 1), E-MVL gradually increases the number of extracted spins $n_{i}(t)$ by scheduling $P_{s}(t)$. The number of extracted spins $n_{i}(t)$, which represents the count of neighboring spins that remain connected and are considered when updating spin $\sigma_{\mathit{i}}$, is calculated on the basis of the current sparsity $P_{s}(t)$ according to the following equation:

\begin{equation}
n_i(t) = \max(1, \lfloor (1 - P_s(t)) \times L_i \rfloor).
\label{eq:definen}
\end{equation}

This formulation ensures that at least 1 spin is extracted even when $P_{s}(t) =$ 1. This extraction number $n_{i}(t)$ determines the number of spins to be considered when updating spin $\sigma_{\mathit{i}}$, directly controlling the degree of sparsification at each iteration. Here, spins with disconnected interactions because of sparsification are called disconnected spins, whereas spins that remain connected and are referenced during the update process are called extracted spins. $L_i$ represents the total number of spins connected to spin $\sigma_{\mathit{i}}$ (including $\sigma_{\mathit{i}}$ itself). These interactions between $\sigma_{\mathit{i}}$ and nonextracted spins are disconnected, and the number of disconnected interactions is denoted by$L_i$– $n_{i}(t)$. Fig.~\ref{fig:Figure1}(b) shows the temporal evolution of $P_{s}(t)$ and the number of extracted spins $n_{i}(t)$. Here, we introduce $P_{s\_init}$ and $P_{s\_fin}$ as the initial and final values of the sparsity schedule, which are set to 1.0 and 0, respectively. Throughout this paper, $t$ denotes the iteration step (or iteration index), and $t_{\mathit{fin}}$ denotes the total number of iterations. As illustrated, $P_{s}(t)$ gradually decreases from 1 to 0 over iterations, whereas $n_{i}(t)$ increases according to Eq. (\ref{eq:definen}) and approaches $L_i$. In the early stages of the solution search, $P_{s}(t)$, which indicates the proportion of sparsification, assumes large values, resulting in small $n_{i}(t)$ values. This emulates the thermal fluctuations in SA and enables spin updates that can escape from local minima, thereby performing a global solution search. As the solution search progresses, $P_{s}(t)$ gradually decreases, causing $n_{i}(t)$ to increase, and the spin states converge to lower-energy states.

For each spin $\sigma_{\mathit{i}}$, the set of extracted spins $M_{i}(t)$ at iteration $t$ is defined as follows:

\begin{equation}
M_i(t) \subset \{1, 2, \dots, L_i\}.
\label{eq:definem}
\end{equation}
The cardinality of this set is $n_{i}(t)$, which is calculated using Eq. (\ref{eq:definen}):
\begin{equation}
|M_i(t)| = n_i(t).
\label{eq:mandn}
\end{equation}

The extracted spins in $M_{i}(t)$ are then used to calculate the internal signal $I_{i}(t+1)$, which determines the next spin state on the basis of the following energy-based decision rule. $M_{i}(t)$ is formed by randomly extracting $n_{i}(t)$ spins from all spins connected to $\sigma_{\mathit{i}}$. This set is defined independently for each spin $\sigma_{\mathit{i}}$ and each iteration t. Each extracted spin is denoted by index $k$ and represented as $\sigma_{\mathit{k}}$. That is, each spin $\sigma_{\mathit{k}}$ corresponding to $k \in M_i(t)$ is considered when updating $\sigma_{\mathit{i}}$. Here, when $k = i$, the external magnetic field $h_{\mathit{i}}$ acting on the spin is considered instead of the spin $\sigma_{\mathit{i}}$ itself. Note that the number of indices $k$ in each extraction set $M_{i}(t)$ equals $n_{i}(t)$, as calculated via Eq. (\ref{eq:definen}). For example, for $L_{i} = 4$, when $P_{s}(0) = 1$, $n_i(0) = \max(1, \lfloor (1 - 1) \times 4 \rfloor) = 1$; thus, exactly 1 spin is extracted. Fig.~\ref{fig:Figure1}(c) shows a detailed visualization of the interaction sparsification mechanism and its temporal evolution throughout the implementation of the E-MVL algorithm. It illustrates the three distinct phases of the ground-state search following the schedule in Fig.~\ref{fig:Figure1}(b), using a 4-spin fully connected Ising model as an example. In the initial phase ($t = 0$), 1 spin is randomly extracted according to Eq. (\ref{eq:definen}), whereas the remaining 3 spins are disconnected. Additionally, the spin update order is set randomly, ensuring search diversity. Thus, in the early stages of solution search, each spin performs local spin updates that minimize energy on the basis of information from only a subset of spins. However, these local energy-minimizing spin updates do not necessarily lead to energy reduction in the original Ising model as a whole. Consequently, spin updates that increase the overall system energy may occur, which represents the local minimum escape effect realized by sparsification. In the intermediate phase $(t = t_1)$, as $P_{s}(t)$ partially decreases, 2 spins are extracted, with 2 spins disconnected. This reduces the probability of energy-increasing spin updates compared with the initial phase, promoting energy convergence. In the final phase $(t = t_{\mathit{fin}})$, when $P_{s}(t)$ reaches 0, all 4 spins are extracted with no disconnected spins. In other words, because all spins are referenced, spin updates that necessarily reduce the overall system energy are performed, leading to energy convergence. Notably, the order of spin updates and the selection of spins to be extracted are performed completely randomly at each step. In the early stages, a high degree of sparsification created significant fluctuations, similar to the thermal fluctuations at high temperatures in SA. By contrast, in the final stages, the system converges to a stable low-energy configuration. From another perspective, E-MVL controls two models—the sparsification ratio and the extracted spin number $n_{i}(t)$—using sparsity as a single parameter. This behavior resembles that of quantum annealing, which controls two models: the quantum term and the classical term. This progressive inclusion of interactions reduces fluctuations in a controlled manner, facilitating the efficient navigation of the energy landscape toward the ground state. Although the basic E-MVL algorithm was previously established \cite{yoshida2022mimicking}, the optimal parameter settings remain unexplored. To evaluate the potential of E-MVL, we conducted systematic parameter optimization by examining different sparsity scheduling approaches and parameter ranges.

Spin state updates are determined by majority voting on the extracted spins obtained through sparsification in Eq. (\ref{eq:definem}). Once the extracted spin set $M_{i}(t)$ is determined, the internal signal $I_{i}(t+1)$ for determining the next state of spin $\sigma_{\mathit{i}}$ is calculated as follows:

\begin{equation}
I_i(t + 1) = 
\begin{cases}
h_i + \displaystyle \sum_{k \in M_i(t), k \neq i} J_{ik} \sigma_k(t), & (i \in M_i(t)) \\
\displaystyle \sum_{k \in M_i(t)} J_{ik} \sigma_k(t). & (i \notin M_i(t))
\end{cases}
\label{eq:cases}
\end{equation}

In Eq. (\ref{eq:cases}), the first term indicates that the external magnetic field$h_{\mathit{i}}$is included in the calculation only when spin $\sigma_{\mathit{i}}$ itself is in the extracted set $M_{i}(t)$. The summation terms represent the contributions from all other extracted spins $k (k \neq i)$ connected to $\sigma_{\mathit{i}}$. Physically, $I_{i}(t+1)$ can be interpreted as the local energy gradient at site $i$, computed using only a subset of the interactions determined by $M_{i}(t)$. This sparsification of interactions at each iteration is a key distinguishing feature of E-MVL compared with other optimization algorithms. Once the internal signal $I_{i}(t+1)$ is calculated, the new spin state $\sigma_{\mathit{i}}(t + 1)$ is determined according to the following rule:

\begin{equation}
\sigma_i(t + 1) = 
\begin{cases}
1, & \text{if } I_i(t + 1) > 0 \\
-1, & \text{else if } I_i(t + 1) < 0 \\
r, & \text{otherwise}
\end{cases}
\label{eq:judge}
\end{equation}

where $r$ is a random variable $r \in \{-1, 1\}$. The "otherwise" condition corresponds to the case when $I_{i}(t+1) = 0$, which represents a balanced state where the extracted interactions do not create a net bias in either direction, and in this case, the spin state is determined by random selection. Fundamentally, the subsequent spin state is determined on the basis of the polarity of the internal signal $I_{i}(t+1)$. As demonstrated in Eqs. (\ref{eq:cases}) and (\ref{eq:judge}), the majority voting logic executes spin updates via the summation of interaction terms and external magnetic field terms. All these computations can be realized using simple logic circuits, rendering them highly amenable to hardware implementation. On the basis of the above mathematical formulations, the overall E-MVL algorithm is presented in Algorithm 1.The spin configuration completed after a specific total number of iterations is considered a solution to the original Ising problem.

\begin{algorithm}[tbp]
\caption{Extraction-type majority voting logic (E-MVL)}
\label{alg:emvl}
Set the schedule of sparsity $P_s(t)$\;
Initialize random spin configuration $\sigma_1, \dots, \sigma_N$\;
\For{$t = 0$ \KwTo $(t_{fin} - 1)$}{
    \For{each spin $\sigma_i$}{
        Calculate $n_i(t)$ by using Eq.~\eqref{eq:definen}\;
        Form $M_i(t)$ by extracting $n_i(t)$ spins\;
        Calculate internal signal $I_i(t+1)$ based on Eq.~\eqref{eq:cases}\;
        Decide new spin state $\sigma_i(t+1)$ following Eq.~\eqref{eq:judge}\;
    }
    Update $P_s(t)$ according to predefined schedule\;
}
Return final spin configuration as solution\;
\end{algorithm}

One of the most significant features of E-MVL is the simplification of operations through sparsification of interspin interactions. In SA, for SK spin-glass problems with $N$ spins, updating each spin requires computing all interactions ($N$ spins), resulting in $O(N^2)$ computational complexity for the entire system. By contrast, E-MVL disconnects a portion of interactions through sparsification, eliminating the need to compute these disconnected interactions, thus requiring less computation than SA does. However, the degree of reduction depends on the sparsity settings, and practical problem solving requires an appropriate sparsity configuration to maintain solution accuracy. The primary advantage of E-MVL lies in its operations consisting primarily of integer additions and simple logic operations, whereas SA requires exponential function evaluations and floating-point arithmetic for Boltzmann distribution-based probability calculations. This simple operational structure makes E-MVL significantly more efficient in dedicated hardware implementations.
Our previous reports \cite{yoshida2022efficient,yoshida2022mimicking} established that $P_{s}(t)$ is equivalent to the temperature parameter $T$ in SA. In SA, convergence to optimal solutions is guaranteed through probabilistic transitions on the basis of the Boltzmann distribution. On the basis of this equivalence, we anticipate that a longer computation time will provide more optimal solutions for E-MVL. Indeed, we confirmed that E-MVL can find approximate solutions to the SK bimodal and achieve more accurate optimization than can SDLs \cite{ito2017prompt,shimada2019calculation,miki2021hybridization} and the highly optimized SA \cite{yoshida2022efficient,yoshida2022mimicking}. However, the impact of different parameter settings on performance is not well understood, and a systematic performance evaluation for larger instances and different coupling distributions is lacking.
This study addresses the comprehensive optimization of E-MVL parameters and provides a systematic performance evaluation against SA algorithms across different scales and coupling distributions. Thus, we investigate the appropriate parameters for $P_{s}(t)$ and iterations in E-MVL to improve both the solution speed and accuracy. Our performance characterization extends beyond basic complexity analysis to include systematic benchmarking against multiple SA implementations. Through these investigations across different coupling distributions (bimodal and Gaussian), we reveal the distinct characteristics between sparsity-based and temperature-based approaches, providing important insights into the sparsity control mechanism.

\subsection{SK model and experimental conditions}

We numerically demonstrated this improvement by solving the Ising problem with $N$ spins and all-to-all connectivity, corresponding to the SK model introduced in studies on spin glasses \cite{aramon2019physics, sherrington1975solvable}. Before evaluating the optimization performance, we first verify the mechanism by which sparsity control enables local minima escape and energy convergence. The SK model serves as a standard benchmark for evaluating optimization algorithms because of their fully connected structure and the computational challenges posed by random, competing interactions. The SK model is particularly relevant, as many combinatorial optimization problems can be transformed into ground-state search problems of Ising spin models \cite{lucas2014ising}. Specifically, we examined both SK-bimodal \cite{aramon2019physics}. For SK-bimodal, the coupling coefficients $J_{\mathit{ij}}$ are drawn from a bimodal distribution taking values from $\{-1, +1\}$ with equal probability. For SK-Gaussian, the coupling coefficients $J_{\mathit{ij}}$ are drawn from a Gaussian distribution with a mean of 0 and standard deviation of 1 and then scaled and quantized to a 10-bit signed integer representation (range: $-512$ to $+511$). In both models, no normalization by system size (e.g.,$1/\sqrt{N}$) is applied to the coupling coefficients, and the external magnetic field is set to $h_{\mathit{i}} = 0$ for all spins. Without the $1/\sqrt{N}$ normalization, the energy changes $\Delta E$ upon a single spin flip scale with both the system size $N$ and the magnitude of the coupling coefficients. For each $N$, we executed 20 different instances 1000 times to collect sufficient statistics. We used solution accuracy as a metric of solution performance and optimized sparsity scheduling to improve accuracy rather than shorten the computation time. The accuracy was calculated by dividing the obtained energy by the ground-state energy. Because obtaining the ground-state energy that can be proven to be the exact solution to this large problem is challenging, we defined the ground-state energy for each instance as the lowest energy found by the Fixstars Amplify Annealing Engine \cite{Fixstars2025} with a maximum execution time (60 [s]), where the total number of iterations was automatically determined by the engine on the basis of the problem size. The algorithms presented in this study, including those for comparison, were executed on the same CPU (AMD Ryzen Threadripper 3990X, 2.90 GHz with 256 GB random-access memory) using the clang version 10.0.0--4ubuntu1 compiler with -O3 optimization. Each trial uses different seed values to ensure statistical independence across experiments.

\section{Results and Discussion}

\subsection{Analysis of the sparsity control mechanism}

\begin{figure}[t]
\centering
\includegraphics[width=\columnwidth]{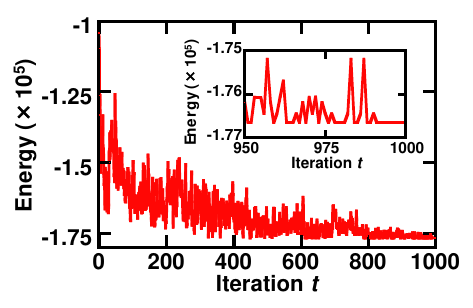}
\caption{Time evolution of energy in a trial that obtained the optimal solution. The inset shows a magnified view of the final stage, demonstrating that continued local minima escape while achieving energy convergence.}
\label{fig:Figure2}
\end{figure}

In the ground-state search of the Ising model, escaping from local minima and achieving energy convergence are critical factors. In this study, we verified whether the sparsity control in E-MVL effectively realizes these factors. Fig.~\ref{fig:Figure2} shows the time evolution of the energy required to obtain the optimal solution when solving a 100-spin SK-Gaussian problem, with an inset presenting a magnified view of the final stage. The time evolution of energy confirms that E-MVL can explore a wide energy range by appropriately incorporating spin updates that increase energy—that is, escape from local minima—through sparsity control. Notably, even in the final stage of the solution search (inset), energy convergence is observed while continuing to perform spin updates to escape local minima.

To verify that the mechanism of sparsity control that realizes such local minima escape and energy convergence is statistically mechanically valid, we examined in detail the energy distributions in the equilibrium state of E-MVL. In our previous study \cite{yoshida2022mimicking}, we demonstrated the statistical mechanical validity of the sparsity control mechanism of E-MVL via the SK-bimodal model. To verify whether these characteristics are preserved in more complex problem instances, we conducted equilibrium energy distribution measurements under identical experimental conditions for the SK-Gaussian model.

\begin{figure*}[t]
\centering
\includegraphics[width=\textwidth]{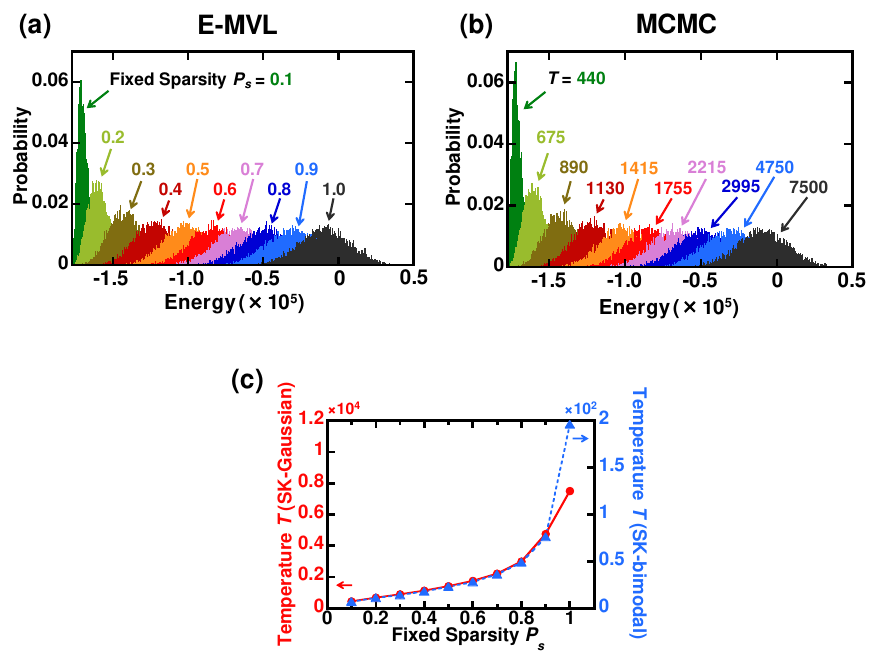}
\caption{Energy distributions at equilibrium for (a) E-MVL and (b) MCMC simulations in SK-Gaussian. Both methods exhibit Boltzmann distributions, demonstrating the probabilistic nature of the energy distribution. (c) Temperature $T$ reproducing the E-MVL equilibrium distributions as a function of fixed sparsity $P_{s}$ for both SK-bimodal and SK-Gaussian, plotted on separate vertical axes.}
\label{fig:Figure3}
\end{figure*}

In these equilibrium measurements, we fixed the sparsity parameter $P_{s}$ at specific constant values and maintained these values unchanged throughout all iterations to observe the energy distribution at each sparsity level. Here, "fixed sparsity" refers to maintaining $P_{s}$ at a constant value throughout all iterations. This is fundamentally different from the time-dependent schedule $P_{s}(t)$ used in ground-state search, where $P_{s}$ gradually decreases from $P_{s\_init}$ to $P_{s\_fin}$. In Fig.~\ref{fig:Figure3}(a) of our previous study \cite{yoshida2022mimicking}, equilibrium energy distributions of E-MVL for the SK-bimodal model were reported when the fixed sparsity $P_{s}$ was varied from 0.1 to 0.9, which shows that at high fixed sparsity values, the energy distributions exhibited large fluctuations, whereas decreasing the fixed sparsity led to reduced variance and a transition toward lower-energy states. In this study, Fig.~\ref{fig:Figure3}(a) shows the equilibrium energy distribution of the E-MVL for the SK-Gaussian model. When the fixed sparsity $P_{s}$ was varied from 0.1 to 1 in the same manner, similar trends to those observed in SK-bimodal were obtained. Specifically, at high fixed sparsity values, the energy distributions exhibited large fluctuations, whereas decreasing fixed sparsity led to reduced variance and a transition of the distribution toward lower-energy states. Note that the plot for $P_{s} = 0$ is excluded because at $P_{s} = 0$, all interactions are considered, leading to convergence to the minimum energy state without establishing an equilibrium state. These results reveal that as $P_{s}$ decreases, the average value of the explored energies decreases, demonstrating that the gradual reduction in $P_{s}$ achieves an efficient transition from initial global exploration to final convergence toward low-energy states.

In contrast, the temperature range required by the MCMC method to reproduce equivalent energy distributions differed substantially depending on the problem type. As shown in Fig.~\ref{fig:Figure3}(b) of our previous study \cite{yoshida2022mimicking}, for the SK-bimodal model, corresponding distributions were obtained in the temperature range of $T = 7$--$75$. However, for the SK-Gaussian model measured in this study, as shown in Fig.~\ref{fig:Figure3}(b), a markedly different temperature range of $T = 440$--$7500$ was required to match the E-MVL equilibrium distributions. Notably, although our previous study \cite{yoshida2022mimicking} did not include measurements at $P_{s} = 1$, the substantial difference in temperature ranges between SK-bimodal ($T = 7$--$75$) and SK-Gaussian ($T = 440$--$7500$) clearly demonstrates the strong dependence of the SA's temperature requirements on the coupling distribution types. To further characterize the equivalence between the sparsity of E-MVL and the temperature of SA, Fig.~\ref{fig:Figure3}(c) shows the temperature values at which MCMC reproduces the equilibrium energy distributions of E-MVL as a function of the fixed sparsity $P_{s}$ for both the SK-bimodal and the SK-Gaussian distributions. Overall, the relationship between $P_{s}$ and $T$ exhibits an approximately exponential dependence for both coupling distributions, which is consistent with previous reports that linearly decreasing $P_{s}$ in E-MVL is expected to produce an effect equivalent to exponentially decreasing $T$ in SA \cite{yoshida2022mimicking}. Notably, despite the substantially different absolute temperature scales between the SK-bimodal and SK-Gaussian distributions, both distributions exhibit nearly identical functional dependences of $T$ on $P_{s}$, further supporting the distribution-independent nature of the sparsity control mechanism. However, a closer inspection reveals that in the range $P_s \leq 0.6$, the relationship becomes approximately linear for both coupling distributions. This distinction is relevant because, as demonstrated in the following sections, the optimal sparsity schedule for E-MVL operates primarily within this linear regime. Notably, at $P_{s} = 1$, where all interactions are disconnected and spin updates are performed on the basis of a single randomly extracted spin, the system exhibits nearly stochastic behavior, resulting in a broad and highly variable energy distribution. Under such conditions, fitting a unique equilibrium temperature becomes ill-defined, as the corresponding temperature value is sensitive to statistical fluctuations. Therefore, the data point at $P_{s} = 1$ should be interpreted with caution. This discrepancy arises because, in the equilibrium state, the spin update probability in SA is expressed as $\min(1, \exp(-\Delta E / T))$, where the energy change $\Delta E$ depends on the coupling coefficient distribution. Moreover, $\Delta E$ also varies with problem size, necessitating careful calibration of the temperature range according to both problem type and size in SA. When $T$ is too large, the energy does not converge to the equilibrium state; when $T$ is too small, the system becomes trapped in local minima. These temperature ranges identified through E-MVL equilibrium analysis suggest that different coupling distributions may benefit substantially from different SA temperature schedules. However, systematic optimization of SA schedules for each problem distribution is beyond the scope of this work. To provide fair and reproducible benchmarking, our SA implementation follows established methodologies from prior work \cite{das2025classical} and uses inverse temperatures $\beta_s = 0.01$ and $\beta_e = 10$ (corresponding to $T = 100$--$0.1$). This configuration serves as our baseline for comparison, representing a broader temperature range than originally recommended for smaller instances.

Importantly, the equilibrium analysis presented here reveals new insights that were not known prior to our investigation. The temperature correspondence obtained through E-MVL equilibrium state measurements—particularly the finding that SK-Gaussian problems correspond to $T = 440$--$7500$—emerges as a novel discovery from this work. This analytical approach, which is based on the equivalence between the sparsity of E-MVL and the temperature of SA, potentially provides a systematic method for estimating appropriate SA temperature ranges for different problem distributions. These findings help explain the performance differences, where the benchmarking results show that SA using the schedule from prior work \cite{das2025classical} has limitations for SK-Gaussian problems at larger scales. The temperature correspondence identified here suggests that different coupling distributions may benefit substantially from different SA schedules, indicating potential directions for SA schedule optimization. However, systematic exploration of such problem-specific SA optimization is beyond the scope of this work, which focuses on demonstrating E-MVL's performance against established SA implementations.

\begin{figure*}[t]
\centering
\includegraphics[width=\textwidth]{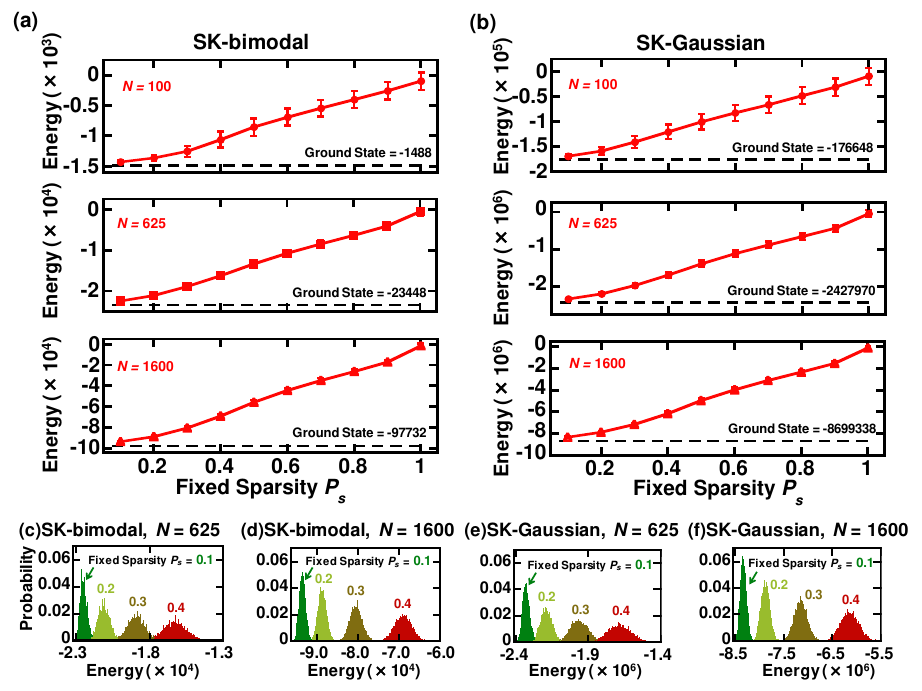}
\caption{ Fixed sparsity-dependent energy characteristics in E-MVL: (a) Average energy and standard deviation versus $P_{s}$ for SK-bimodal and (b) for SK-Gaussian ($N =$ 100, 625, 1600). (c) Energy distributions at $P_{s} =$ 0.1--0.4 for SK-bimodal ($N =$ 625), (d) SK-bimodal ($N =$ 1600), (e) SK-Gaussian ($N$ = 625), and (f) SK-Gaussian ($N =$ 1600) distributions. For energy distributions of $N =$ 100, the results for SK-bimodal are shown in Fig. 3(a) of our previous study \cite{yoshida2022mimicking}, and those for SK-Gaussian are shown in Fig.~\ref{fig:Figure3}(a) of this study.}
\label{fig:Figure4}
\end{figure*}

These results demonstrate a key characteristic of E-MVL: it controls the relative proportion of connected spins rather than directly using energy differences. In SA, the spin update probability depends on $\Delta E$, which varies with both the problem size and the coupling distribution. E-MVL, however, determines spin updates on the basis of the sparsity parameter $P_{s}$, which is independent of the absolute values of these energy changes. To verify this characteristic, we measured the equilibrium energy distributions of the E-MVL across different problem sizes. Figs.~\ref{fig:Figure4}(a) and (b) show the average energy values and their standard deviations obtained from 105 trials in equilibrium states where each fixed sparsity value $P_{s}$ was kept constant for SK-bimodal and SK-Gaussian, respectively. The results are presented for system sizes $N = 100$, 625, and 1600. These results reveal that as $P_{s}$ decreases, the average value of the explored energy decreases. Simultaneously, the standard deviation indicated by the error bars also decreases progressively, demonstrating that the gradual reduction in $P_{s}$ achieves an efficient transition from initial global exploration to final convergence toward low-energy states. Focusing on the region where $P_s \leq 0.4$, where the changes in distribution shape become more pronounced, detailed energy distribution variations are observed. For $N = 100$, the results for SK-bimodal are shown in Fig. 3(a) of our previous study \cite{yoshida2022mimicking}, and those for SK-Gaussian are shown in Fig.~\ref{fig:Figure3}(a) of this study; therefore, these results are omitted from Fig.~\ref{fig:Figure4} to avoid redundancy. Figs.~\ref{fig:Figure4}(c) and (d) show the results for SK-bimodal with $N = 625$ and 1600, respectively, whereas Figs.~\ref{fig:Figure4}(e) and (f) present the results for SK-Gaussian with $N = 625$ and 1600, respectively. These histograms indicate that regardless of the problem size, as $P_{s}$ decreases from 0.4 to 0.1, the variance of the energy distribution decreases markedly, and the distribution peak clearly shifts toward lower energies. In particular, as $P_{s}$ decreases, a sharp peak forms in the low-energy region, confirming that the search range narrows while energy convergence is promoted.

\begin{figure*}[t]
\centering
\includegraphics[width=\textwidth]{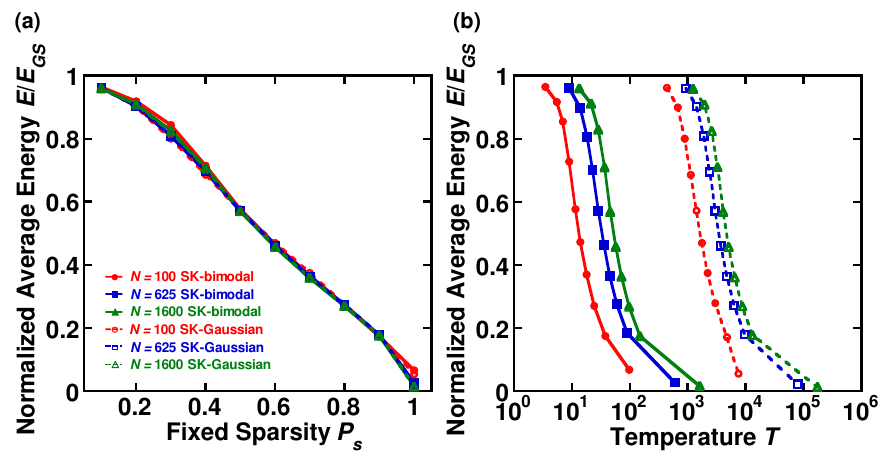}
\caption{Normalized average energy $E/E_{GS}$ in the equilibrium state. (a) $E/E_{GS}$ is plotted as a function of the fixed sparsity parameter $P_{s}$. The plot includes data for both the SK-bimodal and SK-Gaussian models across three system sizes ($N =$ 100, 625, and 1600). All six conditions almost perfectly overlap. (b) $E/E_{GS}$ is plotted as a function of the MCMC temperature $T$ for the same problem types and system sizes. In contrast to (a), the curves exhibit clear separation depending on both the coupling distribution type and the system size.}
\label{fig:Figure5}
\end{figure*}

Notably, the plots of average energy and standard deviation exhibit similar functional forms regardless of the problem type (SK-bimodal or SK-Gaussian) or system size. To quantitatively demonstrate this scale-invariant behavior of E-MVL, Fig.~\ref{fig:Figure5}(a) shows the normalized average energy $E/E_{GS}$ as a function of the fixed sparsity $P_{s}$, where $E_{GS}$ represents the ground-state energy. This normalization enables a comparison of the relative energy exploration range across different problems on a unified scale, even when the absolute values of ground-state energy differ depending on the problem type and size. This figure displays six plots for both the SK-bimodal and SK-Gaussian distributions with $N =$ 100, 625, and 1600, all of which overlap almost perfectly. This result quantitatively demonstrates that, at least within the range of SK models examined, E-MVL can consistently explore a wide range of energy landscapes from high-energy states (low $E/E_{GS}$ region) to low-energy states (high $E/E_{GS}$ region) through consistent sparsity control, regardless of the coupling distribution type or system size. To further substantiate this scale- and problem-invariant property, Fig.~\ref{fig:Figure5}(b) presents the corresponding normalized average energy $E/E_{GS}$ as a function of the MCMC temperature $T$ for the same problem types and sizes. In stark contrast to Fig.~\ref{fig:Figure5}(a), the MCMC curves do not collapse onto a single function. As the system size $N$ increases, the temperature required to achieve a given normalized energy level shifts to substantially higher values. Furthermore, the curves for SK-bimodal and SK-Gaussian remain clearly separated at all system sizes, reflecting the strong dependence of the SA temperature parameter on both the coupling 
distribution and the problem scale. This comparison directly highlights the unique advantage of the sparsity-based control of E-MVL: whereas SA requires problem-specific and size-dependent calibration of the temperature schedule, E-MVL achieves consistent energy landscape exploration through a single sparsity parameter that is largely independent of the coupling distribution and system size. This scale-invariant behavior suggests that the sparsity scheduling strategy optimized for smaller systems can be directly applied to larger problems without modification. Furthermore, this represents a crucial characteristic that suggests the potential for E-MVL to exhibit robust performance across various combinatorial optimization problems beyond the SK models tested here.

\subsection{SA temperature schedule derived from E-MVL equilibrium analysis}

In SA, the temperature schedule is a critical parameter that governs the optimization performance. Isakov et al. \cite{isakov2015optimised} determined that inverse temperatures $\beta_s = 0.1$ and $\beta_e = 3$ were effective for SK problems involving up to 512 spins. Following their methodology, we adopt a broader range of $\beta_s = 0.01$ and $\beta_e = 10$ (corresponding to $T = 100$ to 0.1) as our conventional SA schedule. However, the equilibrium analysis presented in the previous section revealed that the temperature ranges required to reproduce E-MVL equilibrium distributions are substantially greater: $T =$ 7--196 for SK-bimodal and $T =$ 440--7500 for SK-Gaussian (Fig.~\ref{fig:Figure3}(b)). This significant discrepancy between the conventional schedule and the equilibrium-derived temperature ranges motivates the construction of an alternative SA schedule informed by the sparsity–temperature relationship.
\raggedbottom

The sparsity schedule optimization detailed in the Supplementary Information established that the optimal E-MVL sparsity schedule is linear, with $P_{s\_init}$ in the range of 0.2--0.4 and $P_{s\_fin} = 0$. As shown in Fig.~\ref{fig:Figure3}(c), within this optimal range of $P_{s}$, the corresponding temperature $T$ varies approximately linearly with $P_{s}$, enabling a direct translation of the linear sparsity schedule into a linear SA temperature schedule (hereafter referred to as the sparsity-mapped schedule). Note that the temperature at $P_{s} = 0$ cannot be directly measured from the equilibrium analysis because all interactions are fully connected at $P_{s} = 0$ and no equilibrium state is established; instead, the final temperature of the sparsity-mapped schedule is estimated via linear extrapolation of the T($P_{s}$) relationship in the measured range. Because the mapping T($P_{s}$) differs in absolute scale between SK-bimodal and SK-Gaussian while maintaining a similar functional form (as discussed in the context of Fig.~\ref{fig:Figure3}(c)), the resulting sparsity-mapped schedules are distribution specific, reflecting the intrinsic temperature requirements of each coupling distribution. For example, for the SK-Gaussian model with $N =$ 100, the optimal sparsity range $P_{s} =$ 0.4--0 corresponds to a temperature range of approximately $T =$ 890--215, which is substantially different from the conventional schedule range of $T =$ 100--0.1 ($\beta =$ 0.01--10). In the following, we compare SA performance under both conventional and sparsity-mapped schedules to clarify the effect of temperature calibration and to identify the algorithmic contribution of the sparsification mechanism itself.

\begin{figure*}[t]
\centering
\includegraphics[width=\textwidth]{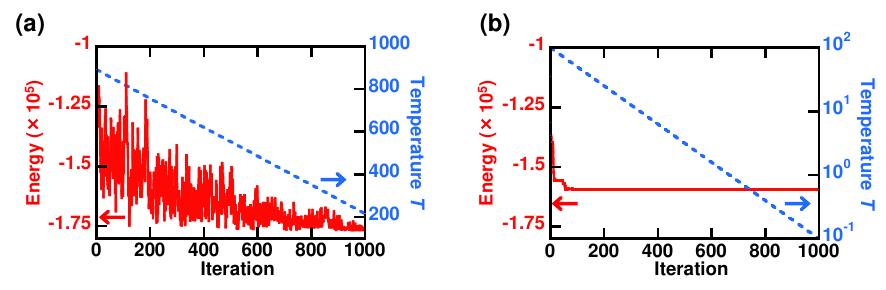}
\caption{Energy evolution (left axis) and temperature schedule (right axis) as a function of iteration for SA applied to a 100-spin SK-Gaussian instance (same as Fig.~\ref{fig:Figure2}). (a) Sparsity-mapped schedule ($T =$ 890 to 215). (b) Conventional schedule ($T =$ 100 to 0.1). Premature freezing is observed in (b), where the energy ceases to decrease well before the final iteration.}
\label{fig:Figure6}
\end{figure*}

Figure~\ref{fig:Figure6} compares the energy evolution and temperature schedule of SA under two different temperature configurations for the same 100-spin SK-Gaussian instance used in Fig.~\ref{fig:Figure2}. In both panels, the horizontal axis represents the iteration number, the left vertical axis shows the energy, and the right vertical axis shows the temperature. Fig.~\ref{fig:Figure6}(a) presents the results under the sparsity-mapped schedule, in which the SA temperature range is derived from the sparsity–temperature mapping T($P_{s}$) established in Fig.~\ref{fig:Figure3}(c). For the SK-Gaussian model with $N =$ 100, the optimal sparsity range $P_{s} =$ 0.4--0 corresponds to a temperature range of approximately $T =$ 890--215. Under this schedule, the energy decreases continuously throughout the iteration process, demonstrating that SA maintains active exploration of the energy landscape without premature freezing. Fig.~\ref{fig:Figure6}(b) presents the corresponding results under the conventional schedule ($T =$ 100 to 0.1, i.e., $\beta =$ 0.01 to 10). As shown in the temperature curve, the conventional schedule rapidly reaches low temperatures in the early stages. Correspondingly, the energy ceases to decrease well before the final iteration, revealing premature freezing of the spin configuration. This premature freezing effectively reduces the useful simulation time, as subsequent iterations no longer contribute to optimization. The comparison between the two panels directly illustrates the origin of the performance difference: the conventional temperature range falls far below the range required for the SK-Gaussian coupling distribution, as revealed by the equilibrium analysis in Fig.~\ref{fig:Figure3}(b), where temperatures of $T =$ 7500 to 440 are required to reproduce the E-MVL equilibrium distributions. By contrast, the sparsity-mapped schedule maintains sufficiently high temperatures during the early stages, providing the thermal fluctuations necessary for global exploration before gradually converging. This result demonstrates that the equilibrium analysis of E-MVL provides a principled and systematic approach for calibrating SA temperature schedules across different coupling distributions. The quantitative performance comparison of E-MVL against SA with both conventional and sparsity-mapped schedules is presented in the following sections using the STT and STS metrics. For each problem size $N$ and coupling distribution, the sparsity-mapped schedule was individually derived from the corresponding $T(P_{s}$) mapping, ensuring that the SA temperature range was appropriately calibrated for each specific problem configuration. 

\subsection{Evaluation of approximate solution}

To quantitatively compare the computational efficiency of E-MVL and SA, we employ the step-to-target (STT) metric \cite{kanao2022simulated,kuroki2024classical}, which evaluates the total number of iterations required to find a target energy (defined as 99\% of the ground-state energy) with 99\% probability (see Supplementary Information for the detailed definitions of the STT and STS metrics). We benchmarked E-MVL against two SA implementations: conventional SA \cite{kirkpatrick1983optimization}, which employs the standard Metropolis criterion, and optimized SA \cite{isakov2015optimised}, which incorporates CPU-specific code-level enhancements (the full algorithms and a comparison with dwave-neal \cite{DWave2025} are provided in the Supplementary Information). For conventional SA, we evaluated performance under both the conventional schedule ($T =$ 100 to 0.1, i.e., $\beta =$ 0.01 to 10) and the sparsity-mapped schedule derived in the previous section to distinguish the effect of temperature calibration from the algorithmic contribution of E-MVL's sparsification mechanism. Optimized SA was evaluated with the conventional schedule to serve as an additional baseline representing a highly optimized SA implementation. For all algorithms, the total number of iterations $t_{\mathit{fin}}$ was optimized to minimize the STT for each problem size $N$ (see Supplementary Information for the optimization details).

\begin{figure*}[t]
\centering
\includegraphics[width=\textwidth]{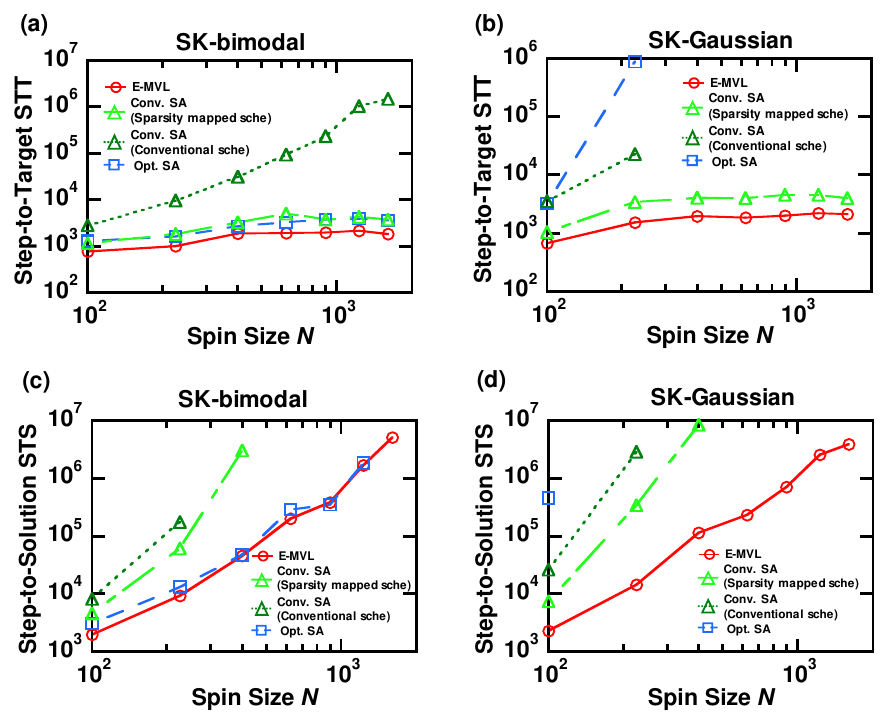}
\caption{ Comprehensive performance comparison for approximate and exact solutions. STT analysis showing computational requirements to obtain 99\% of ground-state energy as a function of $N$ obtained using E-MVL (red solid line), optimized SA (blue dashed line), and conventional SA (green dotted line) with optimal $P_{s}$ and S. (a) Results for SK-bimodal. (b) Results for SK-Gaussian. STS to satisfy absolute ground-state analysis for E-MVL (red solid line), optimized SA (blue dashed line), and conventional SA (green dotted line) with optimal $P_{s}(t)$ and S. (c) Results for SK-bimodal. (d) Results for SK-Gaussian distribution. S is optimized to minimize STS.}
\label{fig:Figure7}
\end{figure*}

Figs.~\ref{fig:Figure7}(a) and (b) show a comparison of the STTs achieved by the E-MVL and SA algorithms for the SK models with 100–1600 spins using the optimized $P_{s}(t)$ and $t_{\mathit{fin}}$ values discussed above. Fig.~\ref{fig:figS2} and Fig.~\ref{fig:figS3} provide the best STT values summarized in Figs.~\ref{fig:Figure7}(a) and (b) for the SK-bimodal and SK-Gaussian models, respectively. For SK-bimodal (Fig.~\ref{fig:Figure7}(a)), both the E-MVL and the optimized SA maintained nearly constant STT values regardless of the problem size N. However, the E-MVL consistently achieved a slightly lower STT than the optimized SA across all values of N, with an approximately 50.7\% reduction at $N =$ 1600. The performance of the conventional SA with the sparsity-mapped schedule achieved performance comparable to that of the optimized SA, indicating that the temperature calibration derived from the E-MVL's equilibrium analysis results in a conventional SA with a performance level comparable to that of highly optimized implementations. By contrast, the conventional SA with the conventional schedule exhibited a quadratic increase in STT as $N$ increased, confirming that the schedule from prior work is suboptimal even for the SK-bimodal distribution at larger scales.

For the SK-Gaussian distribution (Fig.~\ref{fig:Figure7}(b)), the performance differences became substantially more pronounced. The E-MVL method yielded $N$ independent STT values across all tested problem sizes. By contrast, both the optimized SA and the conventional SA with the conventional schedule failed to find approximate solutions within 10,000 iterations for $N \geq 400$. This failure is directly attributable to the premature freezing demonstrated in Fig.~\ref{fig:Figure6}(b): the conventional temperature range ($T =$ 100 to 0.1) falls far below the range required for the SK‒Gaussian coupling distribution ($T =$ 890 to 215), as revealed by the equilibrium analysis. Notably, conventional SA with the sparsity-mapped schedule successfully obtains approximate solutions across all tested problem sizes, including $N \geq 400$, confirming that the premature freezing is effectively resolved through the temperature calibration derived from E-MVL's equilibrium analysis. Nevertheless, E-MVL consistently maintained lower STT values than did conventional SA with the sparsity-mapped schedule across all problem sizes, indicating that the sparsification mechanism itself provides an algorithmic advantage beyond what can be achieved through temperature schedule optimization alone.

The results from both coupling distributions reveal two important findings. First, E-MVL's equilibrium analysis provides a practical methodology for determining effective SA temperature schedules: SA with the sparsity-mapped schedule substantially outperformed SA with the conventional schedule, particularly for SK-Gaussian, where the conventional schedule led to complete failure at larger sizes. Second, even after providing SA with an appropriately calibrated temperature schedule, E-MVL maintained a clear performance advantage, demonstrating that the sparsification mechanism contributes to optimization performance beyond mere temperature calibration. These results indicate that E-MVL makes a dual contribution: it functions as an efficient optimization algorithm on its own right while also providing a principled methodology for improving existing SA-based approaches.

\subsection{Evaluation of the exact solution}

In addition to analyzing the performance of the approximate solutions, we obtained exact solutions via E-MVL and SA to benchmark the advantages of our approach, as shown in Figs.~\ref{fig:Figure7}(c) and (d). Similar to the optimization process for STT, we optimized $P_{s}(t)$ and $t_{\mathit{fin}}$ to minimize the STS for each algorithm and then analyzed the exact solution capabilities. Unlike the evaluation of approximate solutions via STT, we used STS, which represents the total number of iterations required to obtain the exact ground-state energy. Our results reveal significant performance differences across algorithms when both the SK-bimodal and SK-Gaussian models are solved.

For the SK-bimodal instances (Fig.~\ref{fig:Figure7}(c)), E-MVL successfully computed exact solutions for systems containing up to 1600 spins. Among the SA baselines, the optimized SA with the conventional schedule achieved the best SA performance, solving instances up to $N =$ 1225, followed by the conventional SA with the sparsity-mapped schedule up to $N =$ 400. The conventional SA with the conventional schedule was limited to $N =$ 225. For the SK-Gaussian distribution (Fig.~\ref{fig:Figure7}(d)), the performance hierarchy changed: E-MVL maintained its computational capabilities up to 1600 spins, whereas the conventional SA with the sparsity-mapped schedule achieved the best SA performance up to $N =$ 400, and both the optimized SA and the conventional SA with the conventional schedule were restricted to even smaller sizes. These limits arise from our STS calculation methodology, which computes the average STS across all instances for each problem size; if even a single instance cannot be solved to optimality in any trial, the STS value for that entire problem size cannot be determined.

A notable contrast emerges when comparing the approximate and exact solution results. In the approximate solution evaluation (Figs.~\ref{fig:Figure7}(a) and (b)), the conventional SA with the sparsity-mapped schedule achieved a performance closely approaching that of E-MVL, particularly for SK-Gaussian, where it successfully resolved the premature freezing that caused complete failure under the conventional schedule. However, in the exact solution evaluation, the gap between E-MVL and all SA variants widened substantially: E-MVL solved instances up to $N =$ 1600 for both coupling distributions, whereas conventional SA with the sparsity-mapped schedule—the best-performing SA baseline—was limited to $N =$ 400. This widening performance gap suggests that while temperature calibration can largely bridge the efficiency gap for approximate solutions, finding exact ground states demands qualitatively different capabilities that the sparsification mechanism of E-MVL provides more effectively than temperature-based approaches do.

This consistent performance across different coupling distributions within the SK model can be understood through our equilibrium analysis. As demonstrated in Fig.~\ref{fig:Figure5}(a), the normalized energy exploration of E-MVL results in nearly identical functional forms across both the SK-bimodal and the SK-Gaussian methods, as well as across different problem sizes. This scale-invariant property within the SK model framework allows E-MVL to operate with consistent sparsity scheduling ($P_{s\_init} =$ 0.2--0.4) for both coupling distributions tested. In stark contrast, as shown in Fig.~\ref{fig:Figure5}(b), the MCMC curves do not collapse onto a single function, reflecting the strong dependence of the SA's temperature parameter on both the coupling distribution and the problem scale. Even with the sparsity-mapped schedule, which provides problem-specific temperature calibration, SA cannot fully replicate this distribution-independent behavior—a limitation that becomes particularly apparent in the exact solution regime.

However, for the exact ground-state search in Figs.~\ref{fig:Figure7}(c) and (d), the STS of E-MVL is size dependent, increasing with $N$ for both coupling distributions. Whereas E-MVL maintains consistent behavior across both coupling distributions in the SK model, showing similar scaling patterns for SK-bimodal and SK-Gaussian, the emergence of size dependence in exact solution scenarios indicates that achieving problem-size-independent performance for ground-state optimization remains a challenge for future algorithmic development. Nevertheless, E-MVL's ability to use nearly the same sparsity scheduling parameters across different coupling distributions within the SK model, in contrast with SA's requirement for coupling-specific temperature optimization, represents a key algorithmic advantage, explaining the consistent computational efficiency observed across all benchmarking scenarios.

\section{Conclusion}
Here, we reported the performance of a quantum-inspired Ising machine that implements E-MVL and compared it with optimized and conventional SA as well as dwave-neal, another high-performance SA implementation. This study addresses the challenges of systematic parameter optimization for E-MVL and comprehensive performance evaluation against established methods across different coupling distributions. E-MVL disconnects the interactions between spins and searches for the ground state by controlling sparsity $P_{s}(t)$. We confirmed the enhanced solution quality by optimizing $P_{s}(t)$ scheduling based on the solution accuracy. Furthermore, we demonstrated that E-MVL can accelerate the ground-state search with optimal iterations. Through experiments using optimal parameters and metrics independent of hardware implementation, we provided clear evidence of the advantage of E-MVL over all the tested SA implementations on the same CPU to solve the SK model.
 
A key methodological contribution is the equilibrium state analysis, which establishes a quantitative mapping between the sparsity parameter of the E-MVL and the temperature of SA. This mapping revealed that the temperature ranges required for different coupling distributions differ substantially—$T =$ 7--196 for SK-bimodal and $T$ = 440--7500 for SK-Gaussian—far exceeding the conventional schedule ($T =$ 100 to 0.1) adopted from prior work. Translating the optimized sparsity schedule through this mapping yielded the sparsity-mapped SA schedule, which eliminated the premature freezing observed under the conventional schedule and substantially improved SA performance, particularly for SK-Gaussian. This finding demonstrates that E-MVL's equilibrium analysis serves as a practical tool for calibrating SA temperature schedules. Importantly, even with the sparsity-mapped schedule, E-MVL maintained a clear performance advantage. For approximate solutions, SA with the sparsity-mapped schedule closely approached E-MVL's efficiency, yet E-MVL consistently achieved lower STT values. For exact solutions, the gap widened substantially: E-MVL solved instances up to $N =$ 1600 for both coupling distributions, whereas the best SA baseline was limited to $N =$ 400. These results establish a dual contribution of E-MVL: it functions as an efficient optimization algorithm on its own right through the sparsification mechanism while simultaneously providing a principled methodology for improving existing SA-based approaches.

The most significant advantage of E-MVL is its distribution-independent performance. The normalized energy curves confirmed that E-MVL achieves consistent energy landscape exploration through a single sparsity parameter largely independent of the coupling distribution and system size, whereas SA requires problem-specific temperature calibration. In solving SK-Gaussian problems, SA with the conventional schedule failed beyond 400 spins, whereas E-MVL successfully obtained both approximate and optimal solutions up to 1600 spins. FPGA implementation further validated E-MVL's hardware efficiency, with its simple operational structure—primarily integer additions and logic operations—expectedly to yield even greater improvements in ASIC implementations.

The demonstrated scalability, distribution independence, and hardware efficiency position E-MVL as a promising foundation for next-generation dedicated optimization hardware. The sparsification mechanism of E-MVL enables natural extension to other problem structures and offers potential for further performance improvements when combined with other quantum-inspired approaches. These findings contribute to the advancement of efficient combinatorial optimization methodologies and open new possibilities for practical applications in diverse fields requiring large-scale optimization solutions.

\bibliography{reference}

\appendix

\section{Methods}

\subsection{Conventional simulated annealing}

The conventional SA algorithm used in this study implements the standard Metropolis criterion \cite{kirkpatrick1983optimization}. At each time step, a target spin $\sigma_i$ is selected, and the energy change $\Delta E_i$ associated with flipping $\sigma_i$ is calculated from all interactions connected to $\sigma_i$. The flip is accepted if $\Delta E_i \leq 0$ (i.e., the energy decreases or remains unchanged); otherwise, the flip is accepted with probability $\exp(-\Delta E_i / T)$, where T is the current temperature. One sweep consists of $N$ single-spin updates, where the update order is randomized at each sweep. The temperature follows an exponential schedule defined by the inverse temperature:

\begin{equation}
\beta(t) = \beta_s \left( \frac{\beta_e}{\beta_s} \right)^{\frac{t}{t_{fin} - 1}},
\end{equation}

where $\beta_s$ and $\beta_e$ are the initial and final inverse temperatures, respectively, and $t_{fin}$ is the total number of sweeps. In our experiments, we set $T$ = 100 ($\beta_s$ = 0.01) and $T$ = 0.1 ($\beta_e$ = 10). The code is presented in Algorithm 2.

\begin{algorithm}[tbp]
\caption{Conventional Simulated Annealing}
\label{alg:SA_supp}
Set the temperature schedule $T(t)$\;
Initialize random spin configuration $\sigma_1, \dots, \sigma_N$\;
\For{$t = 0$ \KwTo $(t_{fin} - 1)$}{
    \For{each spin $\sigma_i$}{
        Calculate $\Delta E_i$ associated with flipping $\sigma_i$\;
        Generate a uniform random number $u \in [0, 1)$\;
        \If{$\Delta E_i \le 0$ \textbf{or} $u < \exp(-\Delta E_i / T(t))$}{
            Accept the flip: $\sigma_i \leftarrow -\sigma_i$\;
        }
    }
    Update $T(t)$ according to temperature schedule\;
}
Return final spin configuration as solution\;
\end{algorithm}

\subsection{Optimized simulated annealing}
The optimized SA \cite{isakov2015optimised} employs the same Metropolis acceptance criterion as conventional SA but incorporates several CPU-specific implementation-level optimizations to achieve significant computational speedup. The key optimization techniques are summarized below.

\begin{itemize}
    \item \textbf{Forward computation of $\Delta E$.} In conventional SA, the energy change $\Delta E_i$ is recalculated from scratch at every spin update, requiring summation over all interactions connected to $\sigma_i$. In the optimized SA, $\Delta E_i$ is precomputed and stored for all spins at the beginning of the algorithm. When a spin flip of $\sigma_i$ is accepted, only the $\Delta E$ values of spins neighboring $\sigma_i$ are incrementally updated. When a flip is rejected, no update is needed. This eliminates redundant computation, particularly when the acceptance rate is low at low temperatures.
    
    \item \textbf{Fast random number generation.} The optimized SA employs fast pseudorandom number generators such as linear congruential generators. Furthermore, the Metropolis acceptance threshold is precomputed for each temperature step, avoiding repeated floating-point logarithm and exponential evaluations during the spin update loop.
    
    \item \textbf{Loop unrolling with fixed-length loops.} The number of neighboring spins is fixed at compile time, allowing the compiler to perform loop unrolling optimizations. This eliminates loop overhead and enables instruction-level parallelism for the inner loop that computes interaction sums.
\end{itemize}

\subsection{dwave--neal}
dwave-neal \cite{DWave2025} (version 0.6.0) is a high-performance SA implementation developed by D-Wave Systems. It implements the Metropolis‒Hastings algorithm with an exponential temperature schedule. In our experiments, dwave-neal was configured with the same inverse temperature range ($T$ = 100 ($\beta_s$ = 0.01) to $T$ = 0.1 ($\beta_e$ = 10)) and the same number of sweeps as the other SA implementations.

\section{Results}

\subsection{Optimization of the sparsity schedule}

\begin{figure*}[tbp]
\centering
\includegraphics[width=\textwidth]{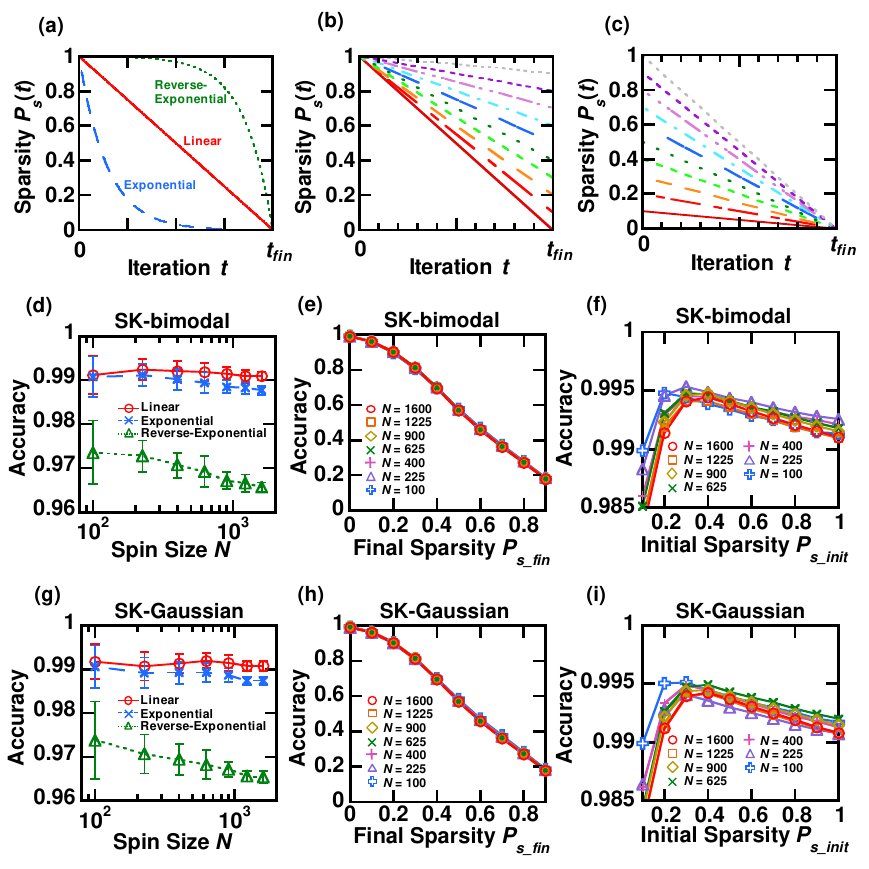}
\caption{Investigation of optimal sparsity scheduling in E-MVL. (a)–(c) illustrate three investigated optimization approaches: (a) Comparison of linear (red solid line), exponential (blue dashed line), and reverse-exponential (green dotted line) scheduling methods; (b) linear schedule with varying final values; and (c) linear schedule with different initial values. (d) and (g) Comparison between three schedules in E-MVL for the Sherrington–Kirkpatrick spin-glass problem with (d) SK-bimodal and (g) SK-Gaussian, corresponding to (a). The red solid line, blue dashed line, and green dotted line correspond to the results obtained via linear, exponential, and reverse-exponential scheduling, respectively. Here, we set initial $P_{s\_init}$ and final $P_{s\_fin}$ at 1 and $10^{-5}$, respectively. Accuracy as a function of $P_{s\_fin}$ for (e) SK-bimodal and (h) SK-Gaussian distributions, corresponding to (b). Accuracy as a function of $P_{s\_init}$ for (f) SK-bimodal and (i) SK-Gaussian distributions, corresponding to (c). To estimate the accuracy of each parameter, we took the average of 20 instances of $10^3$ trials at each $N$.}
\label{fig:figS1}
\end{figure*}

For sparsity schedule optimization, we first choose the best $P_s(t)$ schedule from among the linear, exponential, and reverse-exponential schedules, as conceptually illustrated in Fig.~\ref{fig:figS1}(a), which shows different temporal patterns of sparsity reduction during the solution search process. The linear schedule for $P_s(t)$, denoted as $P_{s\_lin}(t)$, is given by

\begin{equation}
P_{s\_lin}(t) = P_{s\_init} - (P_{s\_init} - P_{s\_fin}) \left( \frac{t}{t_{fin} - 1} \right),
\end{equation}

The exponential schedule for $P_s(t)$, denoted as $P_{s\_exp}(t)$, is given by

\begin{equation}
P_{s\_exp}(t) = P_{s\_init} \left( \frac{P_{s\_fin}}{P_{s\_init}} \right)^{\frac{t}{t_{fin} - 1}},
\end{equation}

The reverse-exponential schedule for $P_s(t)$, denoted as $P_{s\_rev}(t)$, is given by

\begin{equation}
P_{s\_rev}(t) = P_{s\_init} - P_{s\_fin} \left( \frac{P_{s\_init}}{P_{s\_fin}} \right)^{\frac{t}{t_{fin} - 1}},
\end{equation}

where $P_{s\_init}$ represents the initial value of the sparsity schedule $P_s(t)$ (i.e., $t = 0$), and $P_{s\_fin}$ denotes its final value (i.e., $t = t_{fin}$). For this initial comparison, $P_{s\_init}$ and $P_{s\_fin}$ are set to $1$ and $10^{-5}$, respectively, and the total number of iterations $t_{fin}$ is $10^3$. Note that although the asymptote of $P_{s\_rev}(t)$ at $t = t_{fin}$ is $0$ rather than $P_{s\_fin}$, this does not affect the comparison in Fig.~\ref{fig:figS1}, as both values yield the same number of extracted spins $n_i(t_{fin}) = L_i$ when substituted into Eq.~(2). The accuracies of the different schedules as a function of $N$ for both the SK-bimodal and SK-Gaussian cases are shown in Figs.~\ref{fig:figS1}(d) and (g). In both the SK-bimodal and SK-Gaussian cases, the accuracy of the reverse-exponential schedule decreased with increasing $N$, whereas that of the linear and exponential schedules remained constant. Moreover, a slight difference between the linear and exponential accuracies increased as $N$ increased in both models. For comparison, previous studies have shown that when SA is used to search for the ground state of the SK model, an exponential schedule is more effective than linear scheduling because it is more robust against the increasing rate of steps \cite{isakov2015optimised}. By contrast, in E-MVL, the linear schedule resulted in higher accuracy than the exponential schedule for both coupling types because $P_{s\_\mathrm{lin}}(t)$ is equivalent to the exponential schedule of $T$ \cite{yoshida2022mimicking}. The performance advantage of the linear schedule was consistently observed, regardless of the coupling distribution. Therefore, we focused on the linear schedule in the following experiments.

\subsection{Optimization of final sparsity value}

We investigated the final values of the sparsity schedule $P_{s\_fin}$ with a fixed $P_{s\_init}$ of $1$, following the approach illustrated in Fig.~\ref{fig:figS1}(b), which illustrates the concept of varying the final sparsity value while keeping the other parameters constant. We tested $P_{s\_fin}$ values in $0.1$ increments within the range of $0$ to $0.9$ to determine the optimal final sparsity parameter for each problem size $N$. The results of this optimization are shown in Figs.~\ref{fig:figS1}(e) and (h) for the SK-bimodal and SK-Gaussian models, respectively. For both models, the accuracy decreases as the $P_{s\_fin}$ values increase, independent of $N$. This occurs because when $P_{s\_fin}$ is large, sparsification persists even in the final stages of the solution search, causing spin updates to be performed on the basis of only local information. By contrast, when $P_{s\_fin} = 0$, all interactions are considered in the final stages of the solution search, enabling accurate minimization of the entire system energy and achieving the highest accuracy. These consistent results across different coupling schemes indicate that setting $P_{s\_fin}$ to $0$ improved the results.

\subsection{Optimization of initial sparsity value}

Finally, with a fixed linear schedule and $P_{s\_fin}$ at $0$, we optimized the initial value of the sparsity schedule $P_{s\_init}$ for E-MVL to maximize the accuracy, as illustrated in Fig.~\ref{fig:figS1}(c). We tested $P_{s\_init}$ values in $0.1$ increments within the range of $0.1$ to $1$ to determine the optimal initial sparsity parameter for each problem size $N$. The experimental results of this optimization are presented in Figs.~\ref{fig:figS1}(f) and (i) for the SK-bimodal and SK-Gaussian models, respectively. For both models, an excessively large $P_{s\_init}$ may hinder accurate convergence toward the ground state, whereas a small $P_{s\_init}$ immediately converges to a local minimum. The optimal range of $P_{s\_init}$, which offers the highest accuracy, is between $0.2$ and $0.4$ for both coupling types. Furthermore, the optimal $P_{s\_init}$ values remain relatively consistent across different problem sizes $N$ in both cases, with the range of $0.2$--$0.4$ providing good performance regardless of $N$. Each result for a set of scheduling parameters indicates how changes in $P_s(t)$ affect the solution accuracy. The next set of experiments investigated the total number of iterations $t_{fin}$ to optimize the computational time.

\begin{figure*}[tbp]
\centering
\includegraphics[width=\textwidth]{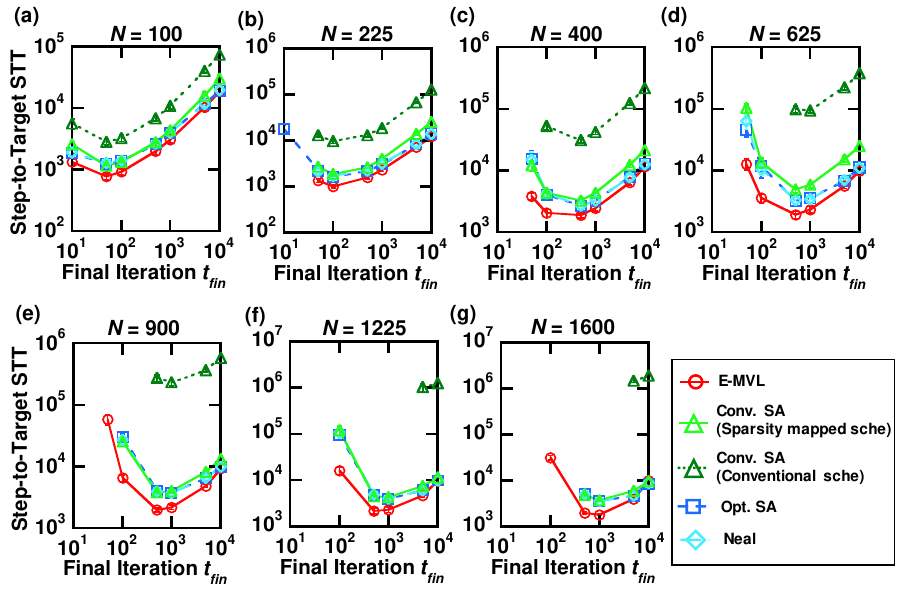}
\caption{Step-to-target (STT) to satisfy 99\% of the ground state as a function of the total number of iterations $t_{fin}$ for each spin size $N$ in SK-bimodal. The lowest STT is achieved at $t_{fin} \in \{50, 100, 500, 1000\}$. The results of the E-MVL and optimized SA methods show similar U-shaped trends.}
\label{fig:figS2}
\end{figure*}

\subsection{Optimization of iteration}

Using the SA algorithms defined above and the sparsity-mapped schedule derived in the main text, we optimized the total number of iterations $t_{fin}$ for each algorithm. To minimize the computational time, we employed the step-to-target (STT) and step-to-solution (STS) metrics \cite{kanao2022simulated, kuroki2024classical}. These metrics evaluate the total number of iterations required to find the target energy with $99\%$ probability, and we utilized them to optimize $t_{fin}$. In general, we employed the time-to-solution (TTS) and time-to-target (TTT) methods to evaluate the computational time of Ising machines \cite{kirkpatrick1983optimization, aramon2019physics, leleu2021scaling}, but these depend on the algorithms, implementations in hardware devices, and coding techniques. To avoid this, STT and STS substitute $t_{fin}$ instead of the computation time per trial in TTS and TTT as follows:

\begin{equation}
\text{STT} = t \frac{\ln(1 - 0.99)}{\ln(1 - P_{target})},
\end{equation}

\begin{equation}
\text{STS} = t \frac{\ln(1 - 0.99)}{\ln(1 - P_{success})},
\end{equation}

where in STT, Ptarget represents the probability of obtaining the target energy, which we define as 99\% of the ground-state energy. By contrast, for STS, $P_{success}$ represents the 
probability of obtaining the exact ground-state energy. Both Ptarget and $P_{success}$ were estimated by dividing the number of successful trials by the total number of trials. In many practical problems, obtaining approximate but practical solutions may be more important than determining the ground-state energy. Therefore, we first examined STT using the target energy instead of the ground-state energy.

\begin{figure*}[htbp]
\centering
\includegraphics[width=\textwidth]{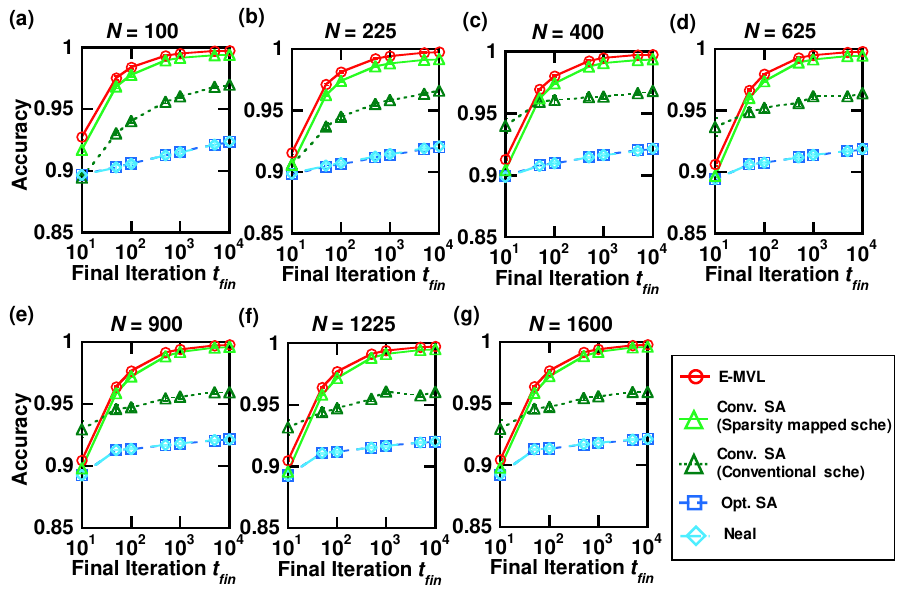}
\caption{Solution accuracy as a function of the total number of iterations $t_{fin}$ for the SK-Gaussian model: (a) $N = 100$, (b) $N = 225$, (c) $N = 400$, (d) $N = 625$, (e) $N = 900$, (f) $N = 1225$, and (g) $N = 1600$. The accuracy is defined as the ratio of the obtained energy to the ground-state energy, averaged over 20 instances with 1,000 trials each. Error bars represent the standard error across instances.}
\label{fig:figS4}
\end{figure*}

\begin{figure*}[t]
\centering
\includegraphics[width=\textwidth]{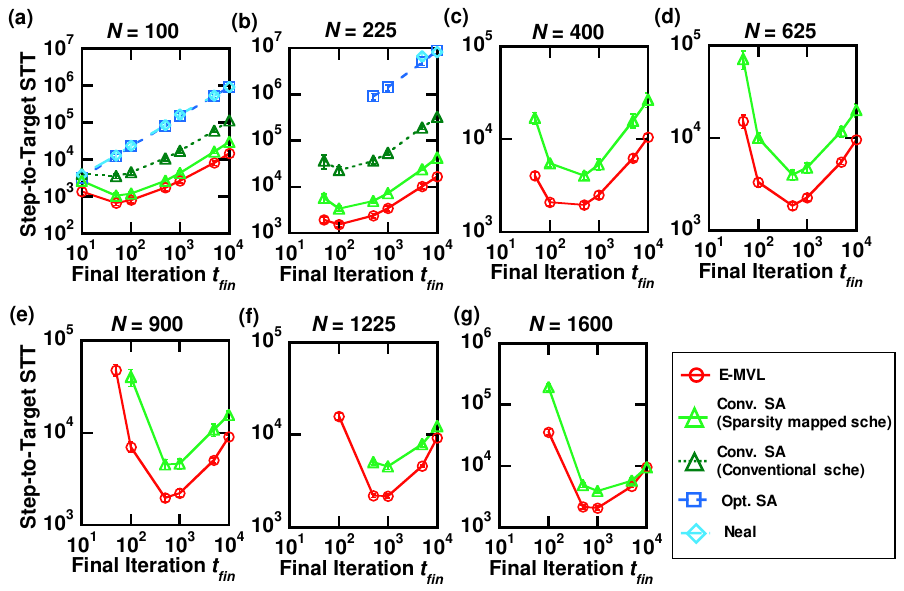}
\caption{STT to satisfy 99\% of the ground state as a function of the total number of iterations $t_{fin}$ for each spin size $N$ in SK-Gaussian. The lowest STT is achieved at $t_{fin} \in \{50, 100, 500, 1000\}$. Similar to the SK-bimodal results, E-MVL exhibits U-shaped trends, with the lowest STT achieved at comparable $t_{fin}$ values. By contrast, both optimized SA and conventional SA fail to find solutions for problem sizes $N \ge 400$ within feasible computational constraints.}
\label{fig:figS3}
\end{figure*}

Figs.~\ref{fig:figS2} and~\ref{fig:figS3} plot the STT results for each $N$ as a function of $t_{fin}$ for the SK-bimodal and SK-Gaussian models, respectively. The comparison includes E-MVL, conventional SA with the conventional schedule ($T = 100$ ($\beta_s = 0.01$) to $T = 0.1$ ($\beta_e = 10$)), conventional SA with the sparsity-mapped schedule, optimized SA with the conventional schedule, and dwave-neal with the conventional schedule. For E-MVL, $P_s(t)$ was optimized to minimize STT for each value of $t_{fin}$. Each $N$ has a minimum STT at a certain $t_{fin}$ in Figs.~\ref{fig:figS2} and~\ref{fig:figS3}. Some plots are not shown because we found that $P_{target}$ = 0 and could not estimate the STT. In both cases, E-MVL achieved lower STT values than the other algorithms, indicating a higher computational efficiency.

For SK-bimodal (Fig.~\ref{fig:figS2}), E-MVL demonstrated comparable or slightly lower STT values than the optimized SA implementation. However, compared with the conventional SA with the conventional schedule, E-MVL achieves a computational efficiency that is up to approximately 100 times greater, as shown in Fig.~\ref{fig:figS2}(g). Conventional SA with the sparsity-mapped schedule achieved performance nearly equivalent to that of optimized SA and dwave-neal, indicating that the temperature calibration derived from E-MVL's equilibrium analysis brings conventional SA to a performance level comparable to highly optimized implementations. Thus, E-MVL offers significant performance improvements, particularly over nonoptimized conventional algorithms. The results of dwave-neal were nearly identical to those of the optimized SA across all problem sizes, suggesting that these two implementations might share similar algorithmic strategies or optimization techniques despite their different origins.

For SK-Gaussian (Fig.~\ref{fig:figS3}), E-MVL exhibited an even more substantial reduction in STT. Notably, for SK-Gaussian instances with $N \ge 400$, both optimized and conventional SA with the conventional schedule fail to find approximate solutions within 10,000 iterations, whereas E-MVL successfully obtains solutions with reasonable STT values. By contrast, conventional SA with the sparsity-mapped schedule successfully obtained approximate solutions across all tested problem sizes, including $N \ge 400$, confirming that the premature freezing caused by the conventional schedule is effectively resolved through the temperature calibration derived from E-MVL's equilibrium analysis. Nevertheless, E-MVL consistently maintained lower STT values than conventional SA with the sparsity-mapped schedule, indicating that the sparsification mechanism itself provides an algorithmic advantage beyond what can be achieved through temperature schedule optimization alone. The pattern of similarity between dwave-neal and optimized SA also persisted in the SK-Gaussian distribution, with both algorithms exhibiting nearly identical performance characteristics. This consistent performance equivalence across distinct problem distributions strongly suggests that dwave-neal and optimized SA may implement fundamentally similar algorithmic approaches despite being developed independently. Furthermore, in contrast to the SK-bimodal case, the conventional SA exhibited lower STT values than the optimized SA for the SK-Gaussian case at smaller problem sizes, suggesting that the optimization techniques in the optimized SA may be specifically calibrated for bimodal distributions and are less effective for Gaussian disorder. These results demonstrate that E-MVL achieves lower STT values than the tested SA implementations across different problem distributions. On the basis of the observed performance equivalence between the optimized SA and dwave-neal across different problem distributions, we henceforth present only the optimized SA results for clarity and conciseness in the following analyses. Note that the performance characteristics shown for the optimized SA are representative of dwave-neal as well.

\begin{figure*}[htbp]
\centering
\includegraphics[width=\textwidth]{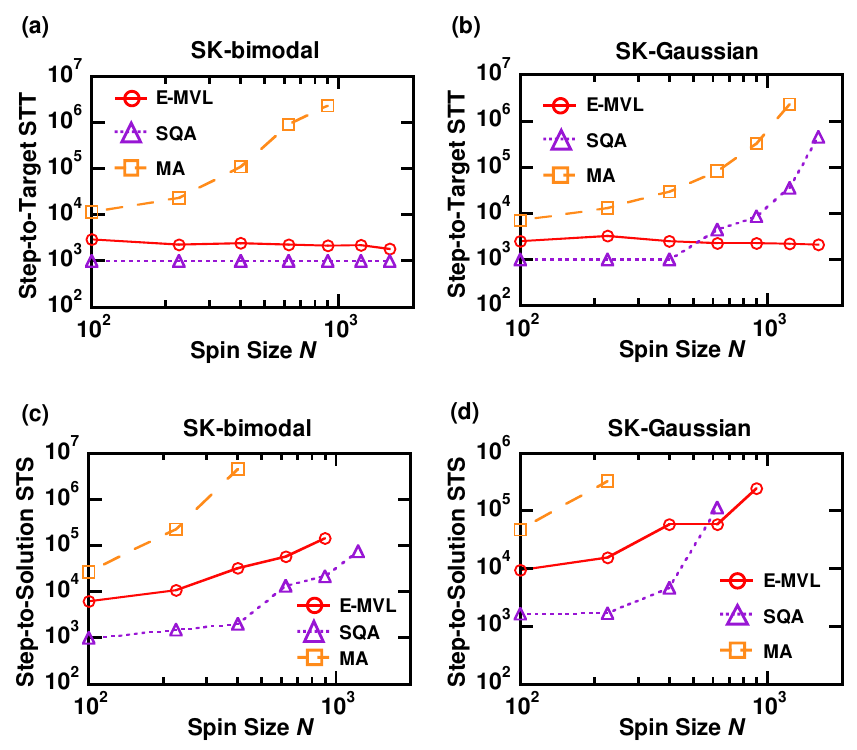}
\caption{Comparison of STT and STS for E-MVL, simulated quantum annealing (SQA), and momentum annealing (MA) as a function of system size $N$. (a) STT for SK-bimodal. (b) STT for SK-Gaussian. (c) STS for SK-bimodal. (d) STS for SK-Gaussian. The total number of iterations was fixed at $t_{fin} = 1000$ for all methods. For each method, algorithm-specific parameters were individually optimized for each problem size and coupling distribution. Error bars represent the standard error from 20 instances with 1000 trials each.}
\label{fig:figS5}
\end{figure*}

\subsection{Solution accuracy analysis for SK-Gaussian}
To further characterize the performance of each algorithm when the target energy is not reached within the iteration budget, Figure~\ref{fig:figS4} presents the solution accuracy as a function of the total number of iterations $t_{fin}$ for the SK-Gaussian model at each problem size $N$. The accuracy is defined as the ratio of the obtained energy to the ground-state energy, averaged over all instances and trials. For $N \ge 400$, where the optimized SA and dwave-neal with the conventional schedule fail to achieve the target energy (99\% of the ground-state energy), these algorithms reach an accuracy of only approximately 0.92 at $t_{fin} = 10,000$, indicating that the conventional temperature schedule leads to premature freezing far from the optimal solution. Conventional SA with the conventional schedule achieves a moderately higher accuracy of approximately 0.96 under the same conditions. By contrast, E-MVL and conventional SA with the sparsity-mapped schedule both exceed an accuracy of 0.99 at $t_{fin} = 1,000$, with E-MVL consistently achieving the highest accuracy among all algorithms. These results quantify the solution quality gap underlying the missing STT data points in Figures~\ref{fig:figS2} and~\ref{fig:figS3} and confirm that the performance advantage of E-MVL extends beyond convergence speed to the quality of solutions obtained at any given iteration budget.

\subsection{Comparison with quantum-inspired methods}
To evaluate E-MVL's performance against established quantum-inspired and advanced classical approaches, we implemented simulated quantum annealing (SQA) \cite{okuyama2017ising} and momentum annealing (MA) \cite{okuyama2019binary} and benchmarked them on the SK model for system sizes up to $N = 1600$. For each method, we optimized the algorithm-specific parameters to ensure fair comparison. The total number of iterations was fixed at $t_{fin} = 1000$ for all methods, including E-MVL, providing a direct comparison under identical computational budgets. This value lies within the near-optimal iteration range for E-MVL as established in the main text (Figs.~\ref{fig:figS2} and~\ref{fig:figS3}). The performance was evaluated using STT and STS metrics, which quantify the intrinsic computational requirements independently of implementation factors. Runtime comparison in seconds is not presented, as CPU-based execution times are sensitive to implementation-specific factors and do not reliably reflect intrinsic algorithmic efficiency.

\begin{table*}[htbp]
\centering
\caption{FPGA resource utilization for E-MVL on Xilinx ZU15EG.}
\label{tab:resource_utilization}
\renewcommand{\arraystretch}{0.9}
\begin{tabular}{lccccc}
\toprule
Model & Spin Size ($N$) & LUT (\%) & Flip-Flop (\%) & BRAM (\%) & DSP (\%) \\
\midrule
\multirow{3}{*}{SK-bimodal}  & 100  & 2892 ($0.85\%$)  & 3428 ($0.50\%$)  & 3.50 ($0.47\%$)  & 2 ($0.06\%$) \\
                             & 625  & 7398 ($2.17\%$)  & 9601 ($1.41\%$)  & 49.50 ($6.65\%$) & 2 ($0.06\%$) \\
                             & 1600 & 15326 ($4.49\%$) & 21987 ($3.22\%$) & 322 ($43.28\%$)  & 2 ($0.06\%$) \\
\midrule
\multirow{3}{*}{SK-Gaussian} & 100  & 1964 ($0.58\%$)  & 2637 ($0.39\%$)  & 6.00 ($0.81\%$)  & 2 ($0.06\%$) \\
                             & 625  & 2224 ($0.65\%$)  & 2752 ($0.40\%$)  & 98.50 ($13.24\%$)& 2 ($0.06\%$) \\
                             & 1600 & 3865 ($1.13\%$)  & 2854 ($0.42\%$)  & 631 ($84.81\%$)  & 2 ($0.06\%$) \\
\bottomrule
\end{tabular}
\end{table*}

\begin{figure*}[htbp]
\centering
\includegraphics[width=\textwidth]{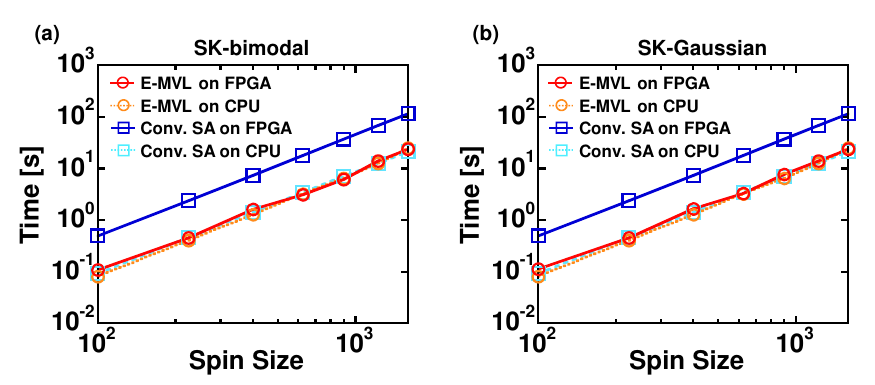}
\caption{Comparison of execution times for the SA and E-MVL algorithms on CPU and FPGA implementations: (a) SK-bimodal problem and (b) SK-Gaussian problem. In the FPGA implementation comparison between E-MVL and SA, E-MVL is 6 times faster than SA for both problem types.}
\label{fig:figS6}
\end{figure*}

Figure~\ref{fig:figS5} shows the STT and STS results as a function of system size $N$ for E-MVL, SQA, and MA. For SK-bimodal (Fig.~\ref{fig:figS5}(a)), both SQA and E-MVL maintained approximately size-independent STT values, whereas MA exhibited increasing STT with system size. SQA achieved the lowest STT values, likely because of the quantum tunneling effects represented through the Suzuki-Trotter decomposition. For SK-Gaussian (Fig.~\ref{fig:figS5}(b)), SQA showed the best performance up to $N = 400$; however, beyond $N = 625$, SQA's STT values increased sharply, and E-MVL achieved the lowest STT among the three methods. MA exhibited even larger STT values for SK-Gaussian. This performance degradation of both SQA and MA for SK-Gaussian at larger system sizes is consistent with the observation that these methods employ temperature-based control: as revealed by the equilibrium analysis in Fig.~\ref{fig:Figure3}, the temperature range required for SK-Gaussian scales substantially with system size, and conventional temperature schedules cannot accommodate the scaling of $\Delta E$ for Gaussian coupling distributions. The STS results (Figs.~\ref{fig:figS5}(c) and (d)) exhibited trends consistent with the STT results.

We note that although runtime comparison was not conducted in this study, MA permits parallel spin updates, making it amenable to GPU implementation with potentially significant speedup. Conversely, SQA achieves high solution accuracy but requires the largest computational resources among the three methods because of the overhead of the Suzuki–Trotter decomposition. These results suggest that applying the sparsity-mapped schedule approach demonstrated for SA in this study to SQA and MA could alleviate their distribution-dependent performance degradation, representing a promising direction for future investigation.

\subsection{FPGA implementation details}
The E-MVL and SA algorithms were implemented on a Xilinx ZU15EG FPGA. Both implementations were evaluated at an operating frequency of 200 MHz. This frequency was determined by the maximum achievable clock rate of our SA implementation; the E-MVL implementation was verified to operate at up to 333 MHz but was set to 200 MHz to ensure a fair comparison.

Table~\ref{tab:resource_utilization} summarizes the FPGA resource utilization for E-MVL across all tested configurations. The resource consumption scales primarily with problem size, with BRAM usage increasing substantially for larger instances because of the storage requirements for coupling coefficients. The DSP utilization remains minimal (0.06\%) across all configurations, reflecting E-MVL's reliance on simple integer arithmetic rather than complex multiply-accumulate operations. The difference in resource profiles between SK-bimodal and SK-Gaussian reflects the nature of the coupling coefficients. For SK-bimodal, where couplings take values from $\{-1, +1\}$, the interaction computation reduces to addition and subtraction operations, allowing the coupling information to be encoded directly into the logic fabric, resulting in higher LUT and Flip-Flop utilization but lower BRAM consumption. For SK-Gaussian, where couplings are represented as 10-bit integers, the coupling coefficients are stored in BRAM and accessed during computation, leading to lower logic utilization but substantially higher BRAM usage, particularly for larger problem sizes (e.g., 84.81\% for $N = 1,600$).

Figure~\ref{fig:figS6} presents the execution times for both algorithms on CPU and FPGA implementations. E-MVL on FPGA achieved execution times comparable to CPU-based implementations despite the more than 10-fold lower clock frequency (FPGA: 200 MHz, CPU: 2.9 GHz). We note that our SA implementation on FPGA did not employ hardware-optimized techniques such as look-up tables for exponential functions or fixed-point approximations of Boltzmann acceptance probabilities, which have been demonstrated to substantially reduce SA's computational overhead in dedicated hardware \cite{nikhar2024all}. Therefore, the SA execution times reported here may not represent the best achievable performance for SA on FPGA, and the speed comparison should be interpreted in the context of our specific implementations.

A particularly noteworthy observation is that FPGA-implemented E-MVL achieved execution times comparable to those of CPU implementations despite the substantial clock frequency disadvantage. This suggests that E-MVL's operational structure—consisting primarily of integer additions and simple logic operations—is well suited for parallel implementation on FPGA, enabling effective parallelization and pipelining that compensates for the lower clock frequency.

\end{document}